# Color Change Effect in an Organic-Inorganic Hybrid Material Based on a Porphyrin Diacid

*Bettina Wagner,[†] Natalie Dehnhardt,[†] Martin Schmid,[†] Benedikt P. Klein,[†] Lukas Ruppenthal,[†] Philipp Müller,[†] Malte Zugermeier,[†] J. Michael Gottfried,[†] Sina Lippert,[‡] Marc-Uwe Halbich,[‡] Arash Rahimi-Iman,[‡] Johanna Heine[†]\**

[†] Department of Chemistry and Material Sciences Center, Philipps-Universität Marburg, Hans-Meerwein-Straße, 35043 Marburg, Germany.

[‡] Department of Physics and Material Sciences Center, Philipps-Universität Marburg, Renthof 5, 35032 Marburg, Germany.

ABSTRACT

Porphyrinic materials show a range of interesting and useful optical and electrical properties. The less well-known sub-class of porphyrin diacids has been used in this work to construct an ionic hybrid organic-inorganic material in combination with a halogenidometalate anion. The resulting compound, [H$_6$TPyP][BiCl$_6$]$_2$ (**1**) (TPyP = tetra(4-pyridyl)porphyrin) has been obtained via a facile solution based synthesis in single crystalline form. The material exhibits a broad photoluminescence emission band between 650 and 850 nm at room temperature. Single crystals of [H$_6$TPyP][BiCl$_6$]$_2$ show a photocurrent in the fA and a much higher dark current in the nA range. They also display an unexpected reversible color change upon wetting with different liquids. This phenomenon has been investigated with optical spectroscopy, SEM, XPS and NEXAFS techniques, showing that a surface-based structural coloration effect is the source of the color change. This stands in contrast to other materials where structural coloration typically has to be introduced through elaborate, multi-step processes or the use of natural templates. Additionally, it underscores the potential of self-assembly of porphyrinic hybrid compounds in the fabrication of materials with unusual optical properties.

INTRODUCTION

Porphyrins feature prominently in many fields of chemistry, ranging from surface modification[1-4] to molecular electronics and solar cells[5-10]. Taking cues from their natural occurrences[11-13] porphyrins and related compounds have been employed in artificial photosynthesis[14,15] as well as a wide range of (photo)catalysis[16-22], sensor[23-25] and medical applications[26-28].[29]

A less well-known subclass are porphyrins featuring an N-H protonation of all four pyrrole rings. These porphyrin diacids can exhibit a distortion of the porphin skeleton[30-32] and a shift in electronic properties towards an acceptor-like behavior. This can be used to construct a variety of organic donor-acceptor materials that show photoconductivity and have been tested for prototype photovoltaic devices.[33-38]

The basic structural and photo-physical features of porphyrin diacids have been well established.[30-32, 39-41] But they have not been extensively used as a building block for crystalline hybrid materials composed of organic and inorganic components, although rare examples exist.[42,43] Organic-inorganic hybrid compounds have seen a rise in scientific interest and breadth of application in the last decades. Typically, they combine the ease of modification for the organic component with the stability and various specific properties of the inorganic component, giving rise to materials with significantly improved or even entirely new properties.[44, 45] Hybrid materials based on free base porphyrins and their metal complexes have been used with success in such diverse areas as heterogeneous catalysis, bactericidal coatings and fuel cell components.[46-50]

Halogenidometalate anions have also become an important building block in hybrid materials, fueled by the discovery and study of hybrid perovskites for solar cell applications.[51-57] For some of the more complex halogenidometalate materials, complex color change phenomena such as photochromism and themochromism have been observed.[58-62] This has prompted us to explore the combination of porphyrins and metal halides into hybrid porphyrin diacid halogenidometalate materials and investigate the ensuing properties.

In this work, we present an example of a compound featuring a porphyrin diacid cation and a strongly interacting chloridobismuthate anion. $[H_6TPyP][BiCl_6]_2$ (**1**) (TPyP = tetra(4-pyridyl)porphyrin) has been obtained in single-crystalline form and characterized regarding its photo-physical properties, including luminescence and photoconductivity. Single crystals of **1** display an unexpected, reversible color change from green to deep blue upon wetting with various liquids. We have used optical spectroscopy, SEM, NEXAFS and XPS techniques to gain insight into this phenomenon.

EXPERIMENTAL METHODS

**General Remarks.** Tetra(4-pyridyl)porphyrin and its HCl salt were synthesized according to literature procedures.[42, 63, 64] All other reagents were used as received from commercial suppliers; reactions were carried out under aerobic conditions. The CHN analysis was performed on an *Elementar* CHN-analyzer. An IR spectrum of **1** was recorded on a *Bruker Tensor 37* FT-IR spectrometer equipped with an ATR-Platinum measuring unit. The spectrum is shown in Figure S5 in the Supporting Information.

**Synthesis.** $Bi_2O_3$ (114 mg, 0.244 mmol) and TPyP (151 mg, 0.244 mmol) were dissolved in 60 mL and 20 mL, respectively, of 2M aqueous HCl solution. The two solutions were mixed, forming a green solution from which blue metallic crystals of **1** started to deposit within a few hours. After one week, the reaction mixture was filtered, the product was washed with 5 mL of cold ethanol and air-dried, yielding 335 mg of **1** (Yield: 90 %).

Data for **1**: Anal. Calcd for $C_{40}H_{38}Bi_2Cl_{12}N_8O_3$, (M = 1522.177 g mol$^{-1}$): C, 31.56; H, 2.52; N, 7.36%. Found: C, 31.33; H, 2.78; N, 7.20%.

**Single-Crystal X-ray Crystallography.** Single crystal X-ray determination was performed on a *Bruker Quest D8* diffractometer with microfocus MoKα radiation and a *Photon 100* (CMOS) detector. The structure was solved using direct methods, refined by full-matrix least-squares techniques and expanded using Fourier techniques, using the Shelx software package[65-67] within the OLEX2 suite[68]. All non-hydrogen atoms were refined anisotropically, unless indicated otherwise. Hydrogen atoms were assigned to idealized geometric positions and included in structure factors calculations. Pictures of the crystal structure were created using DIAMOND.[69] In the refinement of **1**, a number of ISOR restraints had to be used to obtain reasonable anisotropic displacement parameters for the carbon and nitrogen atoms within the porphyrin moiety. Residual electron density corresponding to highly disordered solvent water molecules was treated with a solvent mask as implemented in OLEX2. Crystal data and structure refinement details for **1** have been summarized in Table S1 and selected bond length and bond angles shown in Tables S2 in the Supporting Information. The data for **1** has also been deposited as CCDC 1496256.

**Powder X-ray Diffraction.** Powder patterns, as shown in Figure S3 and Figure S4 in the Supporting Information, were recorded on a *STADI MP* (*STOE Darmstadt*) powder diffractometer, with CuK$\alpha_1$ radiation with $\lambda$= 1.54056 Å at room temperature. The patterns confirm the presence of the phase determined by SCXRD measurement and the absence of any major crystalline by-products. A pattern recorded after heating a sample of **1** at 100°C in vacuum for 6h confirms the stability of the compound.

**Thermal Analysis.** The thermal behavior of **1** was studied by TGA on a *NETZSCH STA 409 C/CD* from 25 °C to 800 °C with a heating rate of 10 °C min$^{-1}$ in a constant flow of 80 mL min$^{-1}$ Ar (see Figure S2 in the Supporting Information). A complex decomposition pattern can be observed with an initial loss of three molecules of solvent water. A detailed analysis is provided in the Supporting Information.

**Scanning Electron Microscopy.** Measurements were performed on a *JEOL JSM-7500F*. Samples were coated with graphite before measurements to ensure sufficient conductivity.

**Optical properties.** Optical absorption spectra were recorded on a *Varian Cary 5000* UV/Vis/NIR spectrometer in the range of 400-800 nm in diffuse reflectance mode employing a Praying Mantis accessory (*Harrick*). For ease of viewing, raw data was transformed from %Reflectance R to Absorbance A according to A = log (1/R).[70]

**Photoluminescence.** The spectra of the photoluminescence measurements were recorded with an Optical Spectrum Analyzer (*ANDO AQ-6315A*). The measurements were performed with a 445 nm continuous-wave laser with a maximum output power of 27 mW.

Time-resolved photoluminescence (TRPL) measurements were performed with a pulsed Titanium-Sapphire Laser (*Spectra Physics, Tsunami*) with a tunable emission wavelength in the range of 700 nm to 1000 nm, a pulse width of 100 fs and a repetition rate of 80 GHz. The light from this laser was frequency doubled by nonlinear optics (*CSK Optronics, Super Tripler 8315*) in order to provide a suitable excitation wavelength (here 445 nm). The photoluminescence was detected by a Streak camera (*Hamamatsu, M5675(s-1)*) with an optimum temporal resolution of 3 ps. In our experiment, a time resolution of about 10-20 ps can be achieved, owing to the detection configuration and settings.

The samples were measured between two quartz plates.

**Resistivity and Photoconductivity.** For photo-electronic measurements, white light from a tungsten lamp was shone on an electrically-contacted sample and the photo-induced current was detected as a function of time, whereas the white light was turned on and off. The emission wavelengths of the lamp ranged from 400nm – 1600nm, with the excitation power of approximately 10μW ($5.1 \cdot 10^{-3}$ W/cm$^2$). Electrical contacts to the sample were established with two tungsten needles, which have a tip diameter of 5 μm. For a sketch of the setup, see the left-side inset of Figure 4. The focused excitation spot on the sample covered approximately an area of hundreds of μm in diameter. An external voltage of +/- 10 V was applied to the sample.

The detection limit of the photocurrent setup was 100 fA. The time resolution for the on-off experiments is 1 s. The photocurrent was detected with a lock-in amplifier (*Stanford Research, SR 850*) after amplification by a current amplifier (*FemtoDLCPA-100*). ). As currents that contribute to the measured signal are modulated by the chopper frequency (180 kHz), the signal from the lock-in amplifier represents the fast response to excitation in contrast to thermal effect.

The *I-V* curves were recorded with the same setup. The current was detected with a *Keithley 617 Programmable Electrometer* in a dark environment. The maximum applied voltage amounted to +/- 15 V.

**NEXAFS and XPS.** Near-edge X-ray Adsorption Fine Structure (NEXAFS) measurements were conducted at the HE-SGM endstation at the *BESSY II* synchrotron facility in Berlin. This dipole-beamline is specifically designed for NEXAFS on light elements (e.g. C, N, O) and allows to measure the photon adsorption indirectly via a Auger/secondary electron detector. The base pressure in the analysis chamber was in the upper $10^{-10}$ mbar range (ultrahigh vacuum, UHV). The samples were introduced via a load-lock into the UHV chamber.

In addition to NEXAFS, X-ray Photoelectron Spectroscopy (XPS) measurements were carried out using Al Kα radiation from a laboratory X-ray source with monochromator, providing a photon energy of 1486 eV, and a *SPECS Phoibos 150* electron analyzer. The base pressure in this setup was in the low $10^{-10}$ mbar range.

RESULTS AND DISCUSSION

**Synthesis.** $[H_6TPyP][BiCl_6]_2$ can be prepared in a straightforward manner starting from either $BiCl_3$ or $Bi_2O_3$. Dissolving the respective bismuth compound together with tetra(4-pyridyl)porphyrin in 2 M hydrochloric acid leads to a deep green solution from which blue-green, metallic crystals of **1** deposit within a few days (see Scheme 1). We have tested a number of different reaction conditions, varying the reagent stoichiometry as well as concentrations of reagents and HCl. We found that, while the synthesis is quite robust, the conditions given above most reliably result in a phase pure product and well-formed single crystals within a short time.

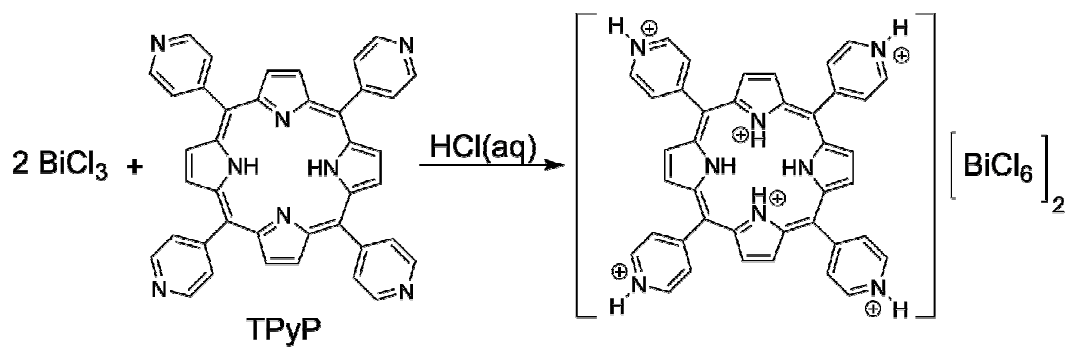

**Scheme 1.** Synthesis of [H$_6$TPyP][BiCl$_6$]$_2$ (**1**).

**Description of the crystal structure.** Compound **1** crystallizes in the monoclinic space group *P*2$_1$/*n* (No. 14). The asymmetric unit consists of one [H$_6$TPyP]$^{6+}$ cation and two [BiCl$_6$]$^{3-}$ anions (see Figure S1 in the Supporting Information). The cation displays a saddle distortion of the porphyrin ring, similar to that found in [H$_6$TPyP]Cl$_6$·H$_2$O[12], well in line with the principles put forth by *Senge* for protonated, *meso*-substituted porphyrin compounds.[30-32]

Isolated [BiCl$_6$]$^{3-}$ anions have been found in a number of hybrid chloridobismuthate compounds. Bi-Cl bond length and Cl-Bi-Cl angles in the respective crystal structures often show a tendency towards distortion when participating in hydrogen bonding interactions.[71-77]

As shown in Figure 1 (a), the [H$_6$TPyP]$^{6+}$ cations form stacks along the *b* axis. Sandwiched in between the cations are highly distorted [BiCl$_6$]$^{3-}$ anions. Similar to what is observed in [H$_6$TPyP]Cl$_6$·H$_2$O, the protonated porphyrin core interacts strongly with the chlorido ligands of the bismuthate complex. Figure 1 (b) gives a side-on view of two adjacent porphyrin/bismuthate stacks and selected interatomic distances. The second type of [BiCl$_6$]$^{3-}$ anion is located outside of these stacks and shows a lesser degree of distortion. Bi-Cl distances found in **1** range from 2.514 to 3.152 Å. This is a more extreme range than typically observed in isolated [BiCl$_6$]$^{3-}$ anions, for example in [C$_6$H$_4$(NH$_3$)$_2$]$_2$ClBiCl$_6$·H$_2$O[74] (Bi-Cl distances from 2.551 to 2.930 Å) but well within the range of bond lengths found in compounds containing multinuclear chloridobismuthate anions such as (BzV)$_5$[Bi$_3$Cl$_{14}$]$_2$ (BzV$^{2+}$ = N,N′-dibenzyl-4,4′-bipyridinium)[78] where Bi-Cl distances range from 2.544 to 3.176 Å. As found in many halogenidometalate compounds, the shortest and longest bonds are *trans* to each other in the [BiCl$_6$]$^{3-}$ anions in compound **1**.[79] In addition to forcing an elongation of Bi-Cl bond length in the [BiCl$_6$]$^{3-}$ anion between the [H$_6$TPyP]$^{6+}$ cations, the interaction also leads to a rather extreme distortion of Cl-Bi-Cl angles, with a maximum angle of 114° between the two *cis* chlorido ligands in closest contact with the porphyrin cations. This degree of distortion is unprecedented in chloridobismuthate compounds, but can be found in other bismuth halogenide compounds such as L·4.88BiBr$_3$ (L = Tetrakis[(4-methylthiophenyl)-ethynyl]-phenylsilane)[80] and [Na$_4$((CH$_3$)$_2$CO)$_{15}$][PtBi$_2$I$_{12}$]$_2$[81] (*cis* X-Bi-X of 114° and 113°, X = Br, I, respectively).

In accordance with their highly charged nature, the [H$_6$TPyP]$^{6+}$ cations show no short intermolecular distances indicative of the pronounced π-π-interactions found in many neutral porphyrin compounds.[82]

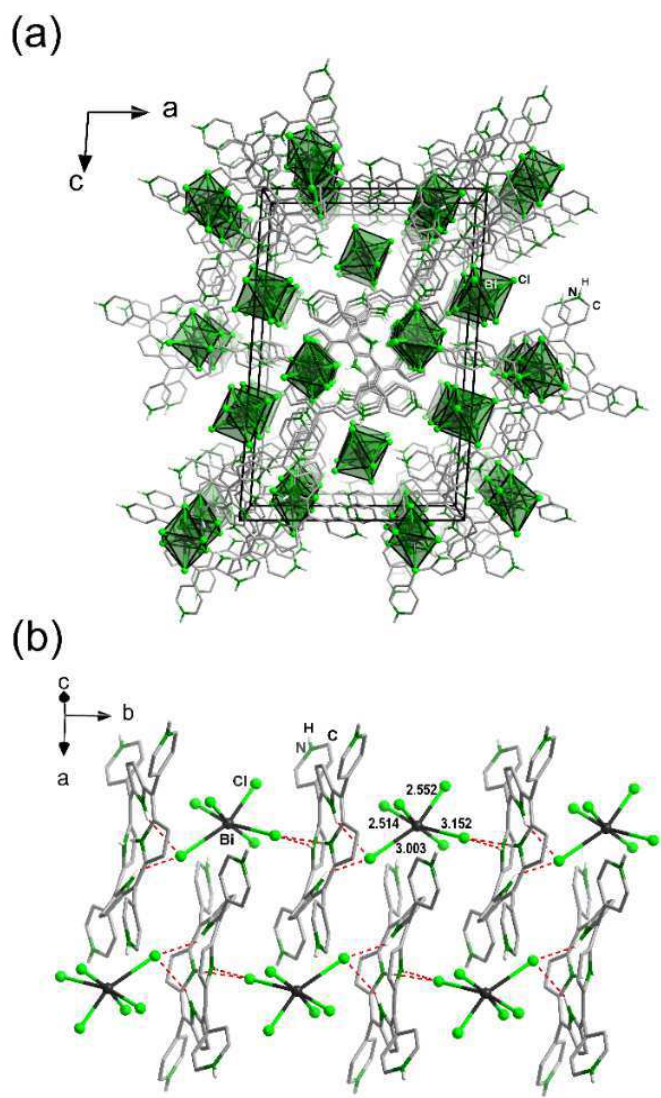

**Figure 1.** (a) Fragment of the crystal structure of **1** showing a view along the *b* axis. (b) Fragment of the crystal structure of **1** showing a side-on view of the stacks of [H$_6$TPyP]$^{6+}$ cations and [BiCl$_6$]$^{3-}$ anions and selected Bi-Cl bond length in Å. Fragmented red lines indicate N-H⋯Cl interactions in the range of 2.4-2.6 Å.

**Optical Properties.** The optical properties of a hybrid compound can typically be traced back either to one of the constituting components or to an interaction of both of them.

The photophysical properties of [BiCl$_6$]$^{3-}$ ions in solution have been studied by *Vogler* and coworkers.[83] They reported absorption bands at 333 and 231 nm, well into the near UV range and in good accordance with the typically colorless appearance of simple chlorobismuthate compounds. Additionally, they found an emission band at 475 nm.

A detailed study of the optical absorption and emission properties of the $[H_6TPyP]^{6+}$ cation in solution has been undertaken by the group of *Scolaro*.[41] They observed a marked influence of the counterion on the absorption properties as well as aggregation behavior of the diacid species. In contrast to this, the emission properties remained nearly unchanged, even when comparing the free base porphyrin with the fully protonated diacids, although a quenching effect upon aggregation was reported. For example, in the case of full protonation of TPyP with trifloroacetic acid, the resulting solution displays emission bands at 665 and 707 nm.

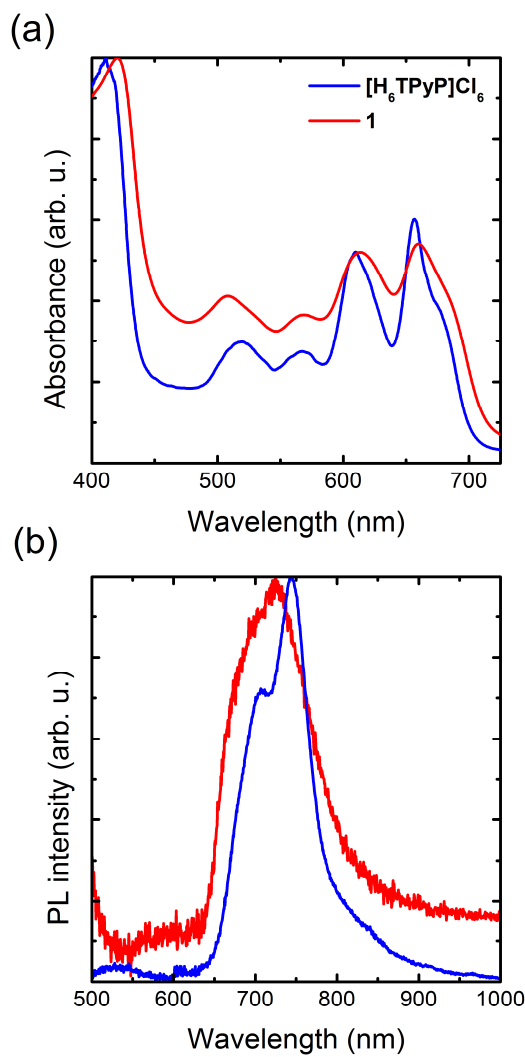

**Figure 2.** (a) UV-Vis spectra recorded in reflectance mode for finely powdered samples of [$H_6$TPyP]$Cl_6$ (blue) and **1** (red). (b) Photoluminescence spectra of [$H_6$TPyP]$Cl_6$ (blue) and **1** (red), respectively.

Figure 2 (a) shows the absorption spectrum, recorded in reflectance mode, for finely powdered samples of [$H_6$TPyP]$Cl_6$ and **1**. Both compounds are similar in the position and intensities of their absorption bands, although a red-shift of about 10 nm can be observed going from [$H_6$TPyP]$Cl_6$ to **1**. The bands also accord fairly well with the results *Scolaro* and coworkers obtained, even though a direct comparison of measurements in solution and the solid state is difficult. Overall, these results suggest that the absorption properties of compound **1** are mainly dominated by the porphyrin diacid cation and that the $[BiCl_6]^{3-}$ anions serve as a template to modify the cation arrangement and thus only indirectly influence the absorption properties of the material.

Photoluminescence measurements of both $[H_6TPyP]Cl_6$ and compound **1** show a broad emission band between 650 and 850 nm. The corresponding spectra of $[H_6TPyP]Cl_6$ and **1** are shown in Figure 2 (b). This indicates that the luminescence properties of **1** are also dominated by the cation. Comparing with *Scolaro's* results for non-aggregated $[H_6TPyP]^{6+}$ cations in solution, we see a red-shift and a broadening of emission bands that result in the observation of a single band instead of two. This is likely due to the transition from essentially isolated $[H_6TPyP]^{6+}$ cations in solution to aggregated ones in the case of solid $[H_6TPyP]Cl_6$ and **1**.

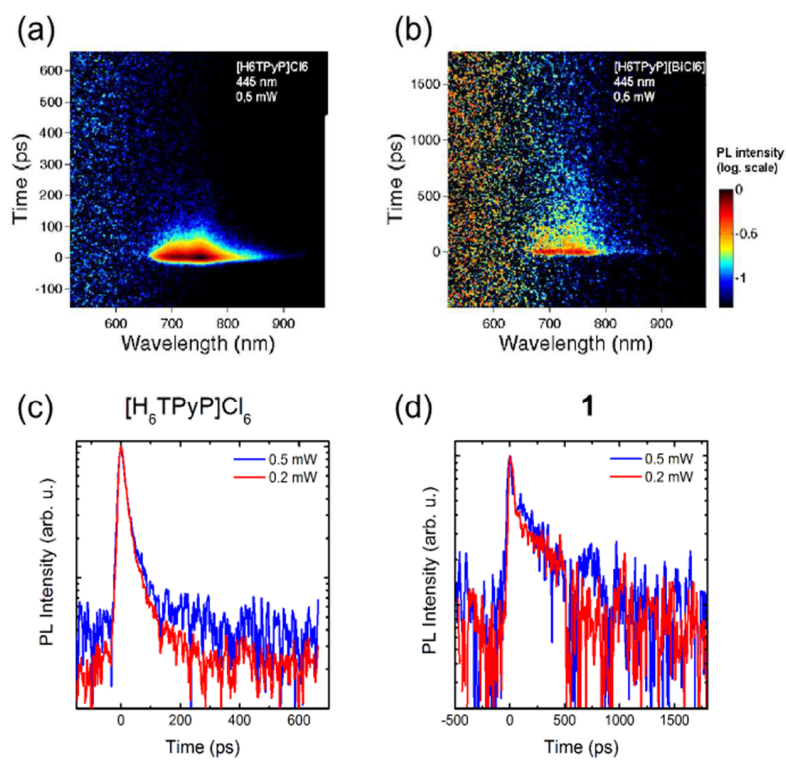

**Figure 3.** (a) Time resolved photoluminescence false color plot of [H$_6$TPyP]Cl$_6$. (b) Time resolved photoluminescence false color plot of **1**. (c) Transient of [H$_6$TPyP]Cl$_6$. (d) Transient of **1**.

Data from the time-resolved photoluminescence (TRPL) are depicted in Figure 3, whereas (a) and (b) represent 2D false color plots for [H$_6$TPyP]Cl$_6$ and **1**, respectively. Figure 3 (c) and (d) show the corresponding transients for both compounds at two different averaged fluences between 0.11 µJ/cm$^2$ and 0.27 µJ/cm$^2$. Here, the photoluminescence of **1** is weaker than the PL of [H$_6$TPyP]Cl$_6$. In addition, it was found that **1** degrades faster with higher laser excitation powers. The transients - obtained from 2D data by integration over the whole wavelength range - indicate that the radiative lifetimes of both compounds exhibit a different behavior. For both, the decay time is governed by two decay channels: one very fast and the other one significantly slower. For [H$_6$TPyP]Cl$_6$, the decay time is approximately 75 ps according to Figure 3 (c). In contrast, the decay time for **1** is approximately 400 ps. The decay times were determined by fitting the data with an exponential function with the equation $y = A * e^{-\frac{t}{\tau}}$, for 30 ps < t < 1 ns. This range corresponds to decay times which can be reliably asserted by our measurement, while no access is given to decay components with time constants shorter than our temporal resolution.

Observations like the weaker luminescence as well as the faster degradation and difference in decay behavior of **1** compared to [H$_6$TPyP]Cl$_6$ can typically be traced back either to energy differences in low-lying anion-porphyrin charge transfer states or a heavy atom effect enhancing intersystem crossing towards triplet states.[40]

**Photoconductivity of Single Crystals.** Porphyrinic materials can show interesting electronic properties including one-dimensional metal-like conduction,[84] semi- and photoconductivity[9]. Often, these properties are found to be critically dependent on several parameters, such as illumination time, the presence of oxygen or other gases or oxidants such as I$_2$.[85-87] *Fukuzumi* and coworkers have demonstrated that photoconductivity can be found for porphyrin diacid compounds co-crystallized with donor molecules.[36] Since chloridobismuthate anions are known to be capable of donating electrons to an acceptor such as methylviologene in photo-activated processes[88,89], we have investigated the photoconductivity and *I-V*-curve of single crystals of **1**.

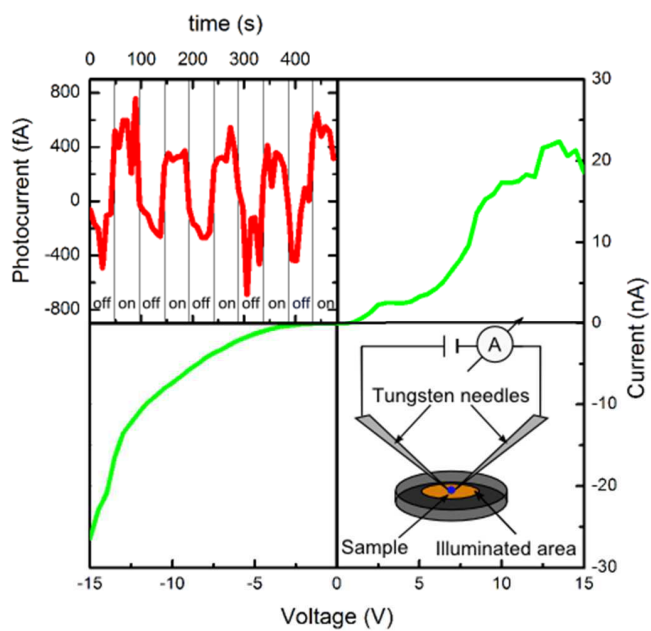

**Figure 4.** Current-voltage (*I-V*) characteristics of **1**. Insets: (left) Time trace of the photocurrent response of **1** upon periodic photoexcitation with a tungsten white-light lamp; (right) Schematic drawing of the photocurrent setup, with the illuminated area indicated by a bright-colored circle, and the sample by a blue dot contacted by needles.

The measured *I-V* curve of **1** is shown in figure 4. The curve shows a clear Schottky contact behavior for positive and negative bias voltages. **1** has a good conductivity which lies between the conductivity of a conductor and an insulator. We attribute the small asymmetry observed in the *I-V*-measurement to the individual contacts between the sample and the tungsten needles. Since the contacts are of a Schottky type, a small difference in the contact quality can lead to a change in the magnitude of current with respect to the direction of flow. The measured on-off photoconductivity over time, i.e. when the exciting light is turned on and off consecutively on a time-scale of a few tens of seconds, is shown in the left-hand inset of Figure 4. A reproducible behavior is recorded over a period of more than 400 s. Here, **1** exhibits a small photocurrent in the range of 600 fA. In comparison to the photocurrent, the dark current of **1** is strong (15 nA at 10 V).

In essence, we observe a conductivity of similar magnitude as in *Fukuzumi's* donor-acceptor salt PNC-TTF (composed of the hydrochloride salt of dodecaphenylporphyrin and tetrathiofulvalene), but only a small change in current upon sample illumination.

**Color Change Effect.** During the course of our investigations, we observed an interesting phenomenon regarding the visible color of single crystals of **1**. While still in the mother liquor, the crystals possess a blue-metallic color. After washing and drying, their color changes to a dull green. To our surprise, the blue color could simply and reversibly be restored by wetting the crystals with different solvents like ethanol or toluene.

Color change in materials is often a complex phenomenon that can depend on external stimuli such as light or oxygen or be triggered by the inclusion and evaporation of solvent molecules. Photochromism has been found for a number of different hybrid chlorido bismuthate compounds, typically with viologene cations. Photochromic effects in these compounds are induced with UV-light, triggering dramatic color changes, for example from colorless to black. These can then be reversed by heating in air.[88-90] Solvatochromic or vapochromic color changes are a ubiquitous phenomenon found in many fields of chemistry and biology,[91-93] as examples in the field of porphyrin compounds[94] and metal halides[95] show.

For the phenomenon at hand, we tested different conditions and found that the change from green to blue depends solely on the wetting of the surface of the crystals and is largely independent of the nature of the solvent, since even very inert liquids such as silicon oil can be used. Single crystal diffraction data, collected from a crystal isolated straight from the mother liquor, corresponds well with the powder patterns collected for samples of **1** dried at 100°C in vacuum (see Figure S4 in the Supporting Information). Together, this shows that a structural change in the material due to the inclusion of solvent molecules is not the cause of the color change. Figure 5 (a) shows the wetting and drying process, recorded in reflectance mode on a UV-Vis spectrometer, as well as photographs of the process (inset). Figure 5 (b) shows the corresponding photoluminescence spectra.

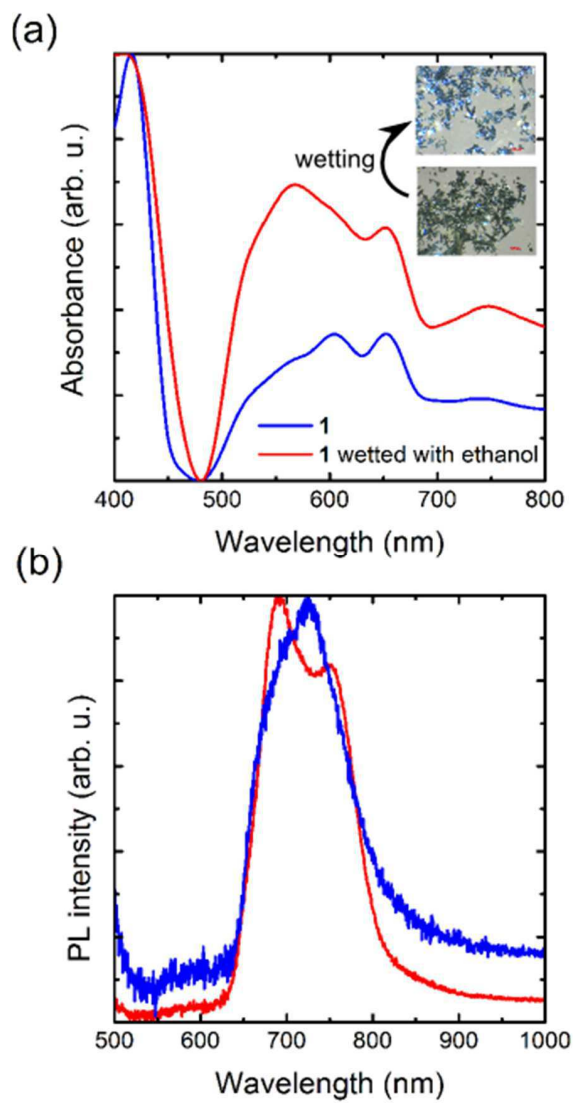

**Figure 5.** (a) UV-Vis spectra recorded in reflectance mode for dry and ethanol-wetted single crystals of **1**. Inset: Photographs of crystals of **1** before (below) and after wetting (above). (b) Photoluminescence of dry (blue) and ethanol-wetted (red) single crystals of **1**, respectively.

We also observed that the color change effect is lost when single crystals of **1** are crushed into a fine powder (see Figure 6a). Due to the robust and unspecific nature of the color change and its dependence on the integrity of the single crystals, we suspected a surface-related structural coloration effect to be the cause of this phenomenon (see Figure 6b for a schematic model). Structural coloration, found in nature for colorful, iridescent butterfly wings or beetles, depends on surface nanostructures generating interference effects and stands in contrast to the light absorption by dyes.[96,97]

Representative photoluminescence spectra for both the wetted as well as the dry crystal only show minor differences. This encourages the assumption that the change in color is a surface related effect, because photoluminescence is normally generated in the bulk.

We performed a SEM investigation on single crystals of **1** and indeed found surface structures in the nanometer range that could be responsible for our observations (Figure 7).

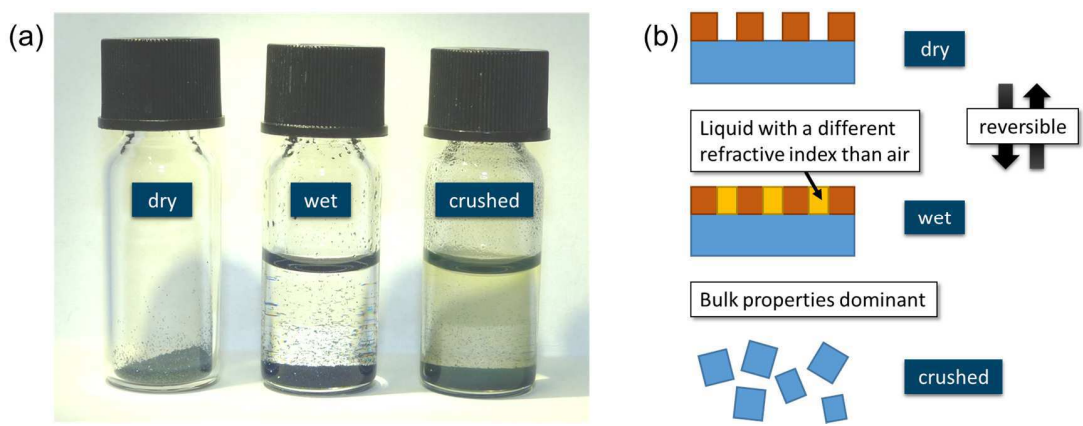

**Figure 6.** (a) Photographs showing, from left to right, dry and toluene-covered single crystals of **1** and the powder resulting from sonication of the crystals in toluene. (b) Schematic model explaining the observed color changes under different conditions. For enhanced clarity, objects in the scheme are not drawn to scale.

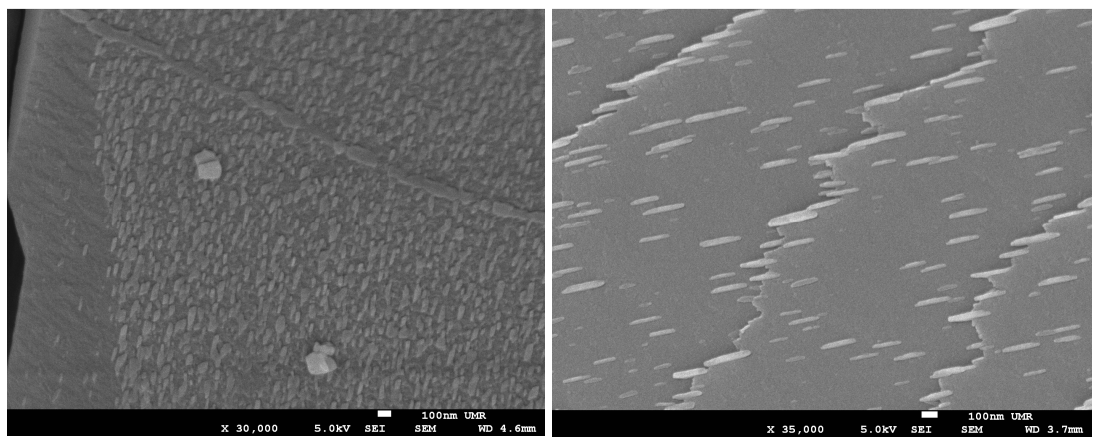

**Figure 7.** SEM images of nanoscale patterns found on single crystals of **1**.

To gain further insight into the composition of these surface features, we studied single crystals of **1** using XPS and NEXAFS techniques.

A survey XP spectrum, shown in Figure S7 in the Supporting Information, confirmed the presence of Bi, Cl and N on the crystals' surface. Yet, a closer examination of the N1s region revealed indications for a missing protonation of the porphyrin units in the near surface region, as shown in Figure 8. The N1s signal comprises three distinct peaks at 389.4 eV (-N=), 399.9 eV ($N_{pyridyl}$), and 401.5 eV (-NH-). This peak pattern is virtually identical to the peak pattern that is observed in monolayers of TPyP[98].

The information depth of XPS is usually defined as three times the photoelectrons' inelastic mean free path in the sample material. This depth range contributes 95% of the XPS signal. Based on the kinetic energy of the N1s photoelectrons of ~1086 eV (resulting from the photon energy of 1486 eV and the binding energy of ~400 eV) one can estimate the information depth, according to an empirical formula[99], to be 3·(0.3)·(1086)0.64 Å = 79 Å = 7.9 nm. Therefore, one can conclude that the first few nanometers of the sample surface in fact contain mainly an uncharged (non-protonated) porphyrin species.

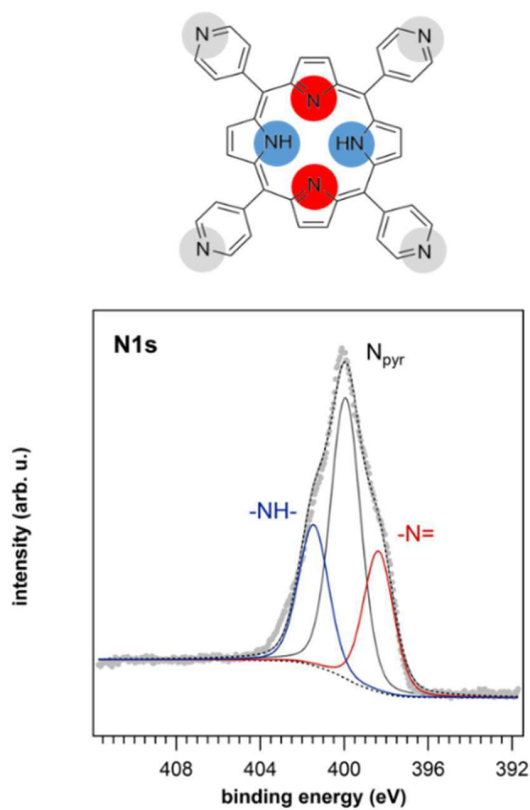

**Figure 8.** X-ray photoelectron spectrum of the N1s region of crystals of **1**. The signal (which originates from a ~7.9 nm thick near-surface layer) shows the spectrum of regular (uncharged) tetrapyridylporphyrin.

To assess whether the missing protonation is limited to the near surface region, NEXAFS spectra were collected in a bulk sensitive and surface sensitive mode (total yield vs. partial yield detection where only electrons with a kinetic energy of more than 150 eV have been counted)[100]. Any difference between spectra recorded in those two measurement modes points to a chemical difference between the chemical composition in the surface and bulk (more accurately, near-surface bulk) region. Indeed, a significant difference between the nitrogen K-edge in total and partial yield detection is observed between 400 and 405 eV (Figure 9). Apart from this difference, both spectra show the expected features for a tetrapyridylporphyrin, i.e. a broad σ* resonance, a π* resonance at 399 eV associated with the pyridyl nitrogen and another π* resonance at 400.5 eV related to the pyrrolic nitrogen (in particular NH) in the cavity of the porphyrin macrocycle.[101]

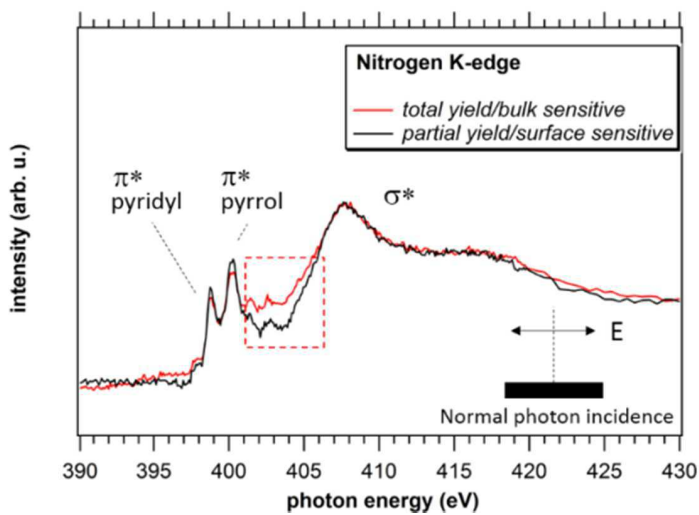

**Figure 9.** Normal-indicence NEXAFS spectra of crystals of **1** measured using partial electron yield (surface sensitive, black curve) and total electron yield (bulk sensitive, red curve). See text for details.

These results strongly suggest that a small amount of a second compound containing bismuth, chloride and a *non-protonated* porphyrin species deposits on the crystals of the bulk material. This could be a coordination compound $[(BiCl_3)_n(H_2TPyP)_m]$, similar to the known family of coordination polymers $[(MX_2)_n(H_2TPyP)_m]$ with M = Mn, Zn, Cd, Hg and X = Cl, Br or I.[102-105] It seems plausible that this phenomenon is induced by a shift in equilibria as the reaction solution is depleted of the starting materials upon product formation, although the large excess of HCl prevents any observable shift in pH value. A schematic model of our current understanding of this process is shown in Figure 10.

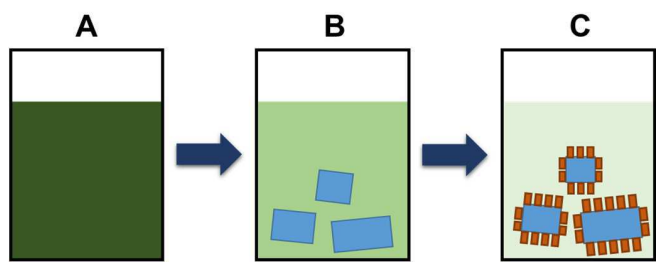

**Figure 10.** Schematic model of the process of nanostructure formation on single crystals of **1**. From the initial solution of starting materials (A) bulk material **1** deposits in single crystalline form (B). Upon depletion of the starting materials, a small amount of a second material forms on the surface of the single crystals of **1**, generating the nanoscale pattern responsible for the color change (C). For enhanced clarity, objects in the scheme are not drawn to scale.

Preliminary studies with differently substituted porphyrin diacids and other bismuth halogenides suggest that the phenomenon is not unique to compound **1**, but can be observed in related compounds as well. This points to a more general aggregation and surface deposition process in solution. Aggregation phenomena are well known for porphyrin compounds, especially for charged species[105,107] and it has been demonstrated that halogenide ions can play a significant role in these processes.[41,108] This has already been used to construct micro- and nanoscale aggregates.[109,110] The exact nature of the occurrence of structural coloration in our materials will be the subject of further in-depth study. However, it appears feasible that a deposition technique can be developed to create thin films or coatings containing structured layers of different hybrid materials. Additionally, the emergence of structural coloration in single crystals of **1** through a solution-based aggregation process stands in stark contrast to the typically very elaborate protocols involving either multiple steps or the use of naturally occurring templates that are necessary to produce this effect artificially in other materials.[111]

CONCLUSIONS

In summary, the new hybrid compound [$H_6$TPyP][BiCl$_6$]$_2$ has been prepared via a facile route and characterized in depth concerning its optical properties. These properties can be mainly attributed to the porphyrin diacid, although the [BiCl$_6$]$^{3-}$ anions appear to enhance non-emissive relaxation pathways in comparison with the simple hydrochloride salt [$H_6$TPyP]Cl$_6$. In photoconductivity measurements, compound **1** displays a significant dark current – similar in magnitude as the photocurrent found for known photoconductive porphyrin diacid compounds – and a comparatively small photocurrent. The color change phenomenon that can be observed upon wetting single crystals of [$H_6$TPyP][BiCl$_6$]$_2$ can be traced back to a surface-based structural coloration effect, as SEM, XPS and NEXAFS measurements show. These also confirm that the surface species is another hybrid compound based on Bi, Cl and non-protonated porphyrin TPyP, possibly a coordination polymer. These results suggest that organic-inorganic hybrid compounds based on halogenidometalates and porphyrin diacids are a promising class of materials for creating structured thin films with interesting optical properties.

ACKNOWLEDGMENT

This work was supported by the Deutsche Forschungsgemeinschaft within the framework of SFB 1083 and the Wissenschaftliches Zentrum für Materialwissenschaften (WZMW) of the Philipps-Universität Marburg. J. H. thanks Prof. Stefanie Dehnen, Philipps-Universität Marburg, for her constant support.

ASSOCIATED CONTENT

**Supporting Information.**

Crystallographic details, thermal analysis data, powder diffraction patterns, IR spectrum, additional UV/Vis spectra and additional XPS data.

AUTHOR INFORMATION

**Corresponding Author**

*J. Heine: johanna.heine@chemie.uni-marburg.de

**Author Contributions**

The manuscript was written through contributions of all authors. All authors have given approval to the final version of the manuscript.

**Notes**

The authors declare no competing financial interest.

**Supplementary Information on**

# Color Change Effect in an Organic-Inorganic Hybrid Material Based on a Porphyrin Diacid

*Bettina Wagner,[†] Natalie Dehnhardt,[†] Martin Schmid,[†] Benedikt P. Klein,[†] Lukas Ruppenthal,[†] Philipp Müller,[†] Malte Zugermeier,[†] J. Michael Gottfried,[†] Sina Lippert,[‡] Marc-Uwe Halbich,[‡] Arash Rahimi-Iman,[‡] Johanna Heine[†]\**

[†] Department of Chemistry and Material Sciences Center, Philipps-Universität Marburg, Hans-Meerwein-Straße, 35043 Marburg, Germany.

[‡] Department of Physics and Material Sciences Center, Philipps-Universität Marburg, Renthof 5, 35032 Marburg, Germany.

## Crystallographic details

**Table S1:** Crystallographic data for **1**.

| | 1 |
|---|---|
| Empirical formula | $C_{40}H_{32}Bi_2Cl_{12}N_8$ |
| Formula weight /g·mol$^{-1}$ | 1468.09 |
| Crystal color and shape | blue block |
| Crystal size | 0.21 x 0.09 x 0.09 |
| Crystal system | monoclinic |
| Space group | $P2_1/n$ |
| $a$ /Å | 18.6824(10) |
| $b$ /Å | 9.7571(5) |
| $c$ /Å | 27.8985(10) |
| $\alpha$ /° | 90 |
| $\beta$ /° | 93.921(2) |
| $\gamma$ /° | 90 |
| $V$ /Å$^3$ | 5073.6(4) |
| $Z$ | 4 |
| $\rho_{calc}$ /g·cm$^{-3}$ | 1.922 |
| $\mu(Mo_{K\alpha})$ /mm$^{-1}$ | 7.598 |
| measurement temp. /K | 100 |
| Absorption correction type | multi-scan |
| Min/max transmission | 0.6149/0.7452 |
| $2\Theta$ range /° | 4.72-50.44 |
| No. of measured reflections | 109025 |
| No. of independent reflections | 9202 |
| $R$(int) | 0.1052 |
| No. of indep. reflections ($I > 2\sigma(I)$) | 7613 |
| No. of parameters | 559 |
| $R_1$ ($I > 2\sigma(I)$) | 0.0376 |
| $wR_2$ (all data) | 0.0717 |
| $S$ (all data) | 1.027 |
| $\Delta\rho_{max}$, $\Delta\rho_{min}$ /e· Å$^{-3}$ | 1.177/-1.172 |

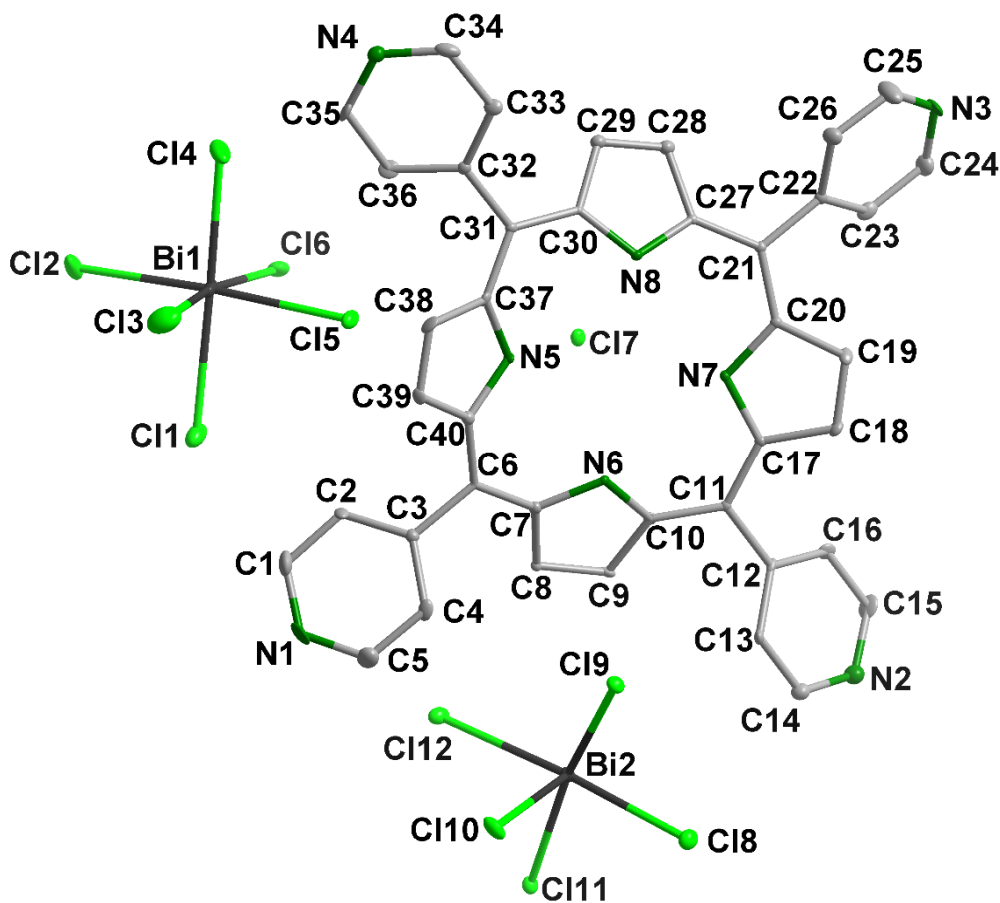

**Figure S1:** Asymmetric unit of **1**, ellipsoids at 50% probability.

**Table S2:** Selected interatomic distances (in Å) and angles (in °) in **1**.

| | | | | | |
|---|---|---|---|---|---|
| Bi2 Cl9 | 3.0033(15) | Cl8 Bi2 Cl9 | 79.65(5) | Cl4 Bi1 Cl6 | 94.07(6) |
| Bi2 Cl12 | 2.7961(16) | Cl8 Bi2 Cl12 | 171.29(5) | Cl4 Bi1 Cl1 | 177.13(6) |
| Bi2 Cl8 | 2.5898(15) | Cl11 Bi2 Cl9 | 160.06(5) | Cl4 Bi1 Cl2 | 90.30(6) |
| Bi2 Cl11 | 2.5521(15) | Cl11 Bi2 Cl12 | 83.35(5) | Cl2 Bi1 Cl5 | 173.37(6) |
| Bi2 Cl10 | 2.5139(16) | Cl11 Bi2 Cl8 | 89.26(5) | Cl2 Bi1 Cl6 | 94.23(6) |
| Bi1 Cl5 | 2.7101(15) | Cl10 Bi2 Cl9 | 77.87(5) | Cl2 Bi1 Cl1 | 89.42(6) |
| Bi1 Cl6 | 2.9288(19) | Cl10 Bi2 Cl12 | 88.58(5) | Cl3 Bi1 Cl5 | 92.03(6) |
| Bi1 Cl1 | 2.7850(18) | Cl10 Bi2 Cl8 | 95.63(5) | Cl3 Bi1 Cl6 | 171.69(6) |
| Bi1 Cl4 | 2.6146(18) | Cl10 Bi2 Cl11 | 86.88(5) | Cl3 Bi1 Cl1 | 86.39(6) |
| Bi1 Cl2 | 2.6938(18) | Cl5 Bi1 Cl6 | 80.86(5) | Cl3 Bi1 Cl4 | 90.77(7) |
| Bi1 Cl3 | 2.593(2) | Cl5 Bi1 Cl1 | 86.04(5) | Cl3 Bi1 Cl2 | 92.50(7) |
| | | Cl1 Bi1 Cl6 | 88.80(5) | | |
| Cl12 Bi2 Cl9 | 108.74(4) | Cl4 Bi1 Cl5 | 94.48(5) | | |

**Thermal analysis**

The thermal behavior of **1** (16.6 mg), was studied by simultaneous DTA/TG on a *NETZSCH STA 409 C/CD* in from 25 °C to 800 °C with a heating rate of 10 °C min$^{-1}$ in a constant flow of 80 ml min$^{-1}$ Ar.

The first mass loss of 3% starting at 115°C corresponds to the loss of three molecules of water per formula unit. Thus, isolated **1** has formula of [H$_6$TPyP][BiCl$_6$]$_2$· 3 H$_2$O, as confirmed by CHN analysis. Starting at 230°C a three step mass loss sets in, indicating a complex decomposition. The high residual mass of 65% points to a bismuth-containing species as one of the decomposition products.

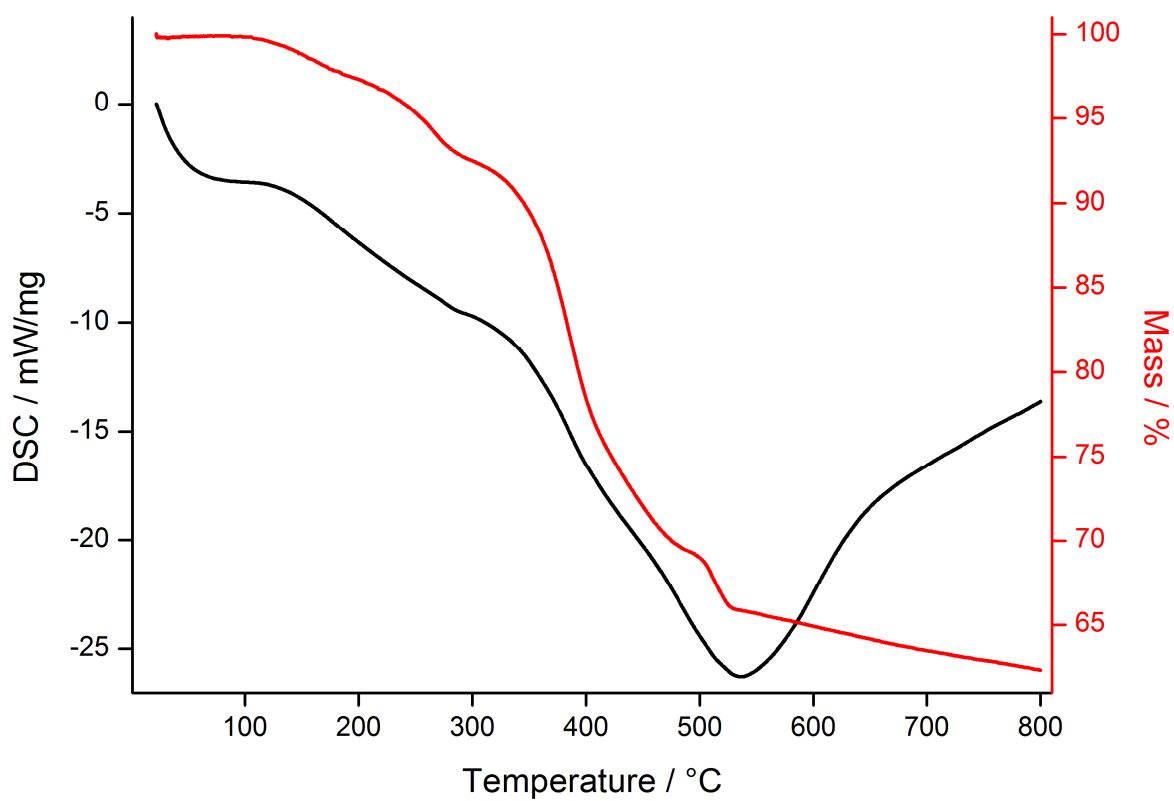

**Figure S2.** Simultaneous DTA/TG of **1**.

**Powder diffraction**

Powder patterns were recorded on a *STADI MP* (*STOE* Darmstadt) powder diffractometer, with CuK$_{\alpha 1}$ radiation with λ= 1.54056 Å at room temperature. The patterns confirm the presence of the phase determined by SCXRD measurement and the absence of any major crystalline by-products. Additionally, a sample was heated to 100°C for 6h under vacuum. The powder pattern recorded afterwards shows that the compound is stable under these conditions.

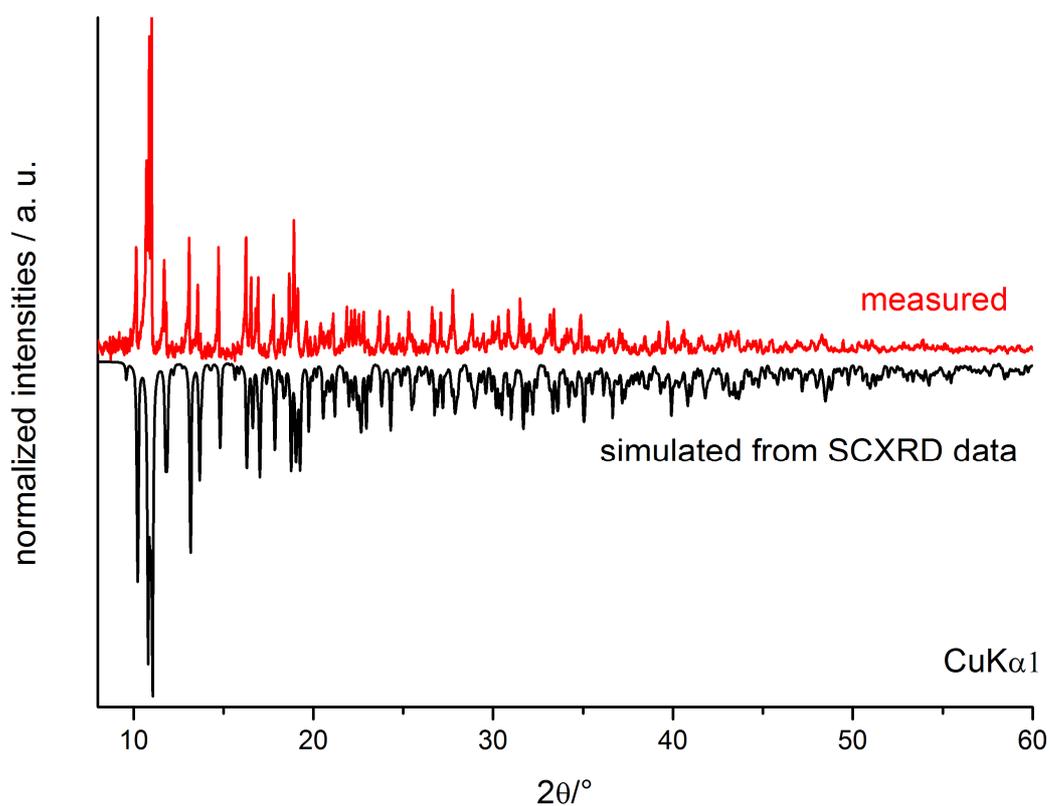

**Figure S3.** Powder diffraction pattern of compound **1**.

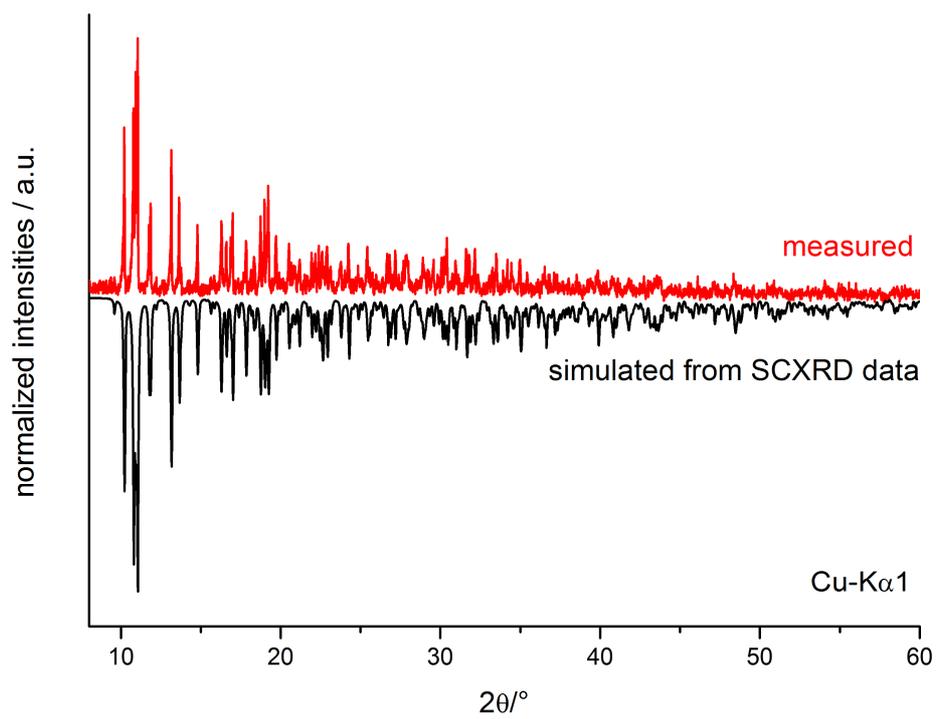

**Figure S4.** Powder diffraction pattern of compound **1** after heating to 100°C for 6h under vacuum.

**IR spectroscopy**

An IR spectrum of **1** was recorded on a *Bruker Tensor 37* FT-IR spectrometer equipped with an ATR-Platinum measuring unit.

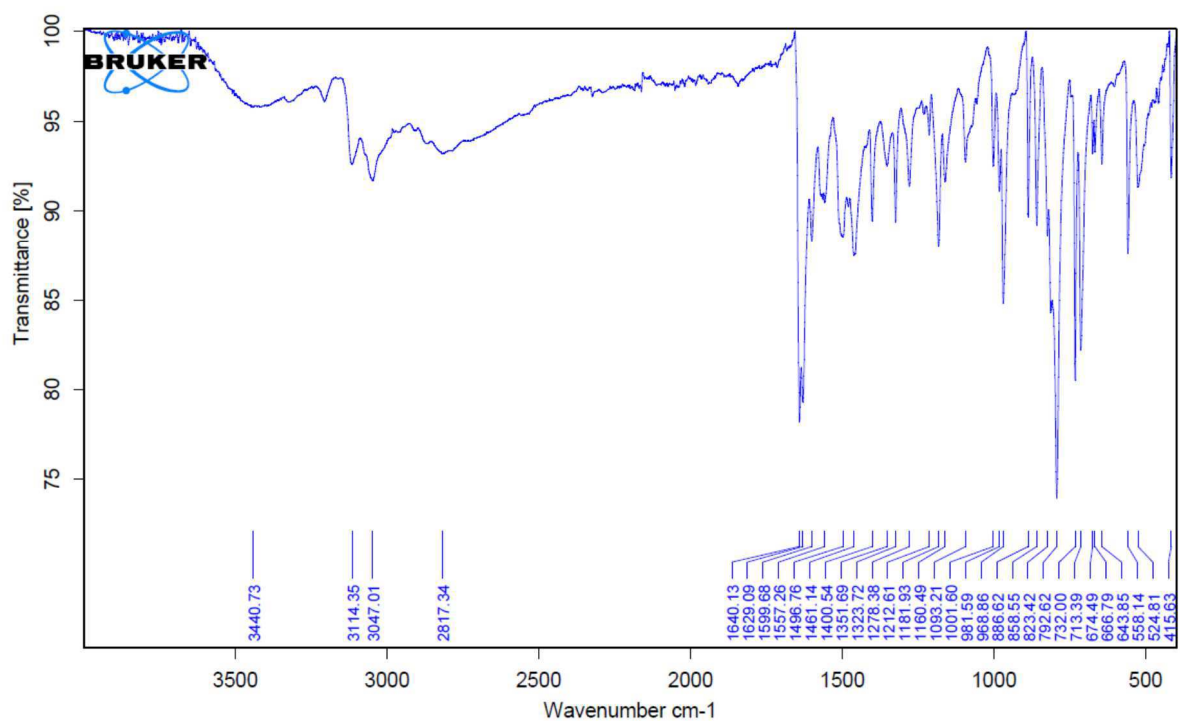

**Figure S5.** IR spectrum of compound **1**.

**Optical properties**

Optical absorption spectra were recorded on a *Varian Cary 5000* UV/Vis/NIR spectrometer in the range of 800-200 nm in diffuse reflectance mode employing a Praying Mantis accessory (*Harrick*).

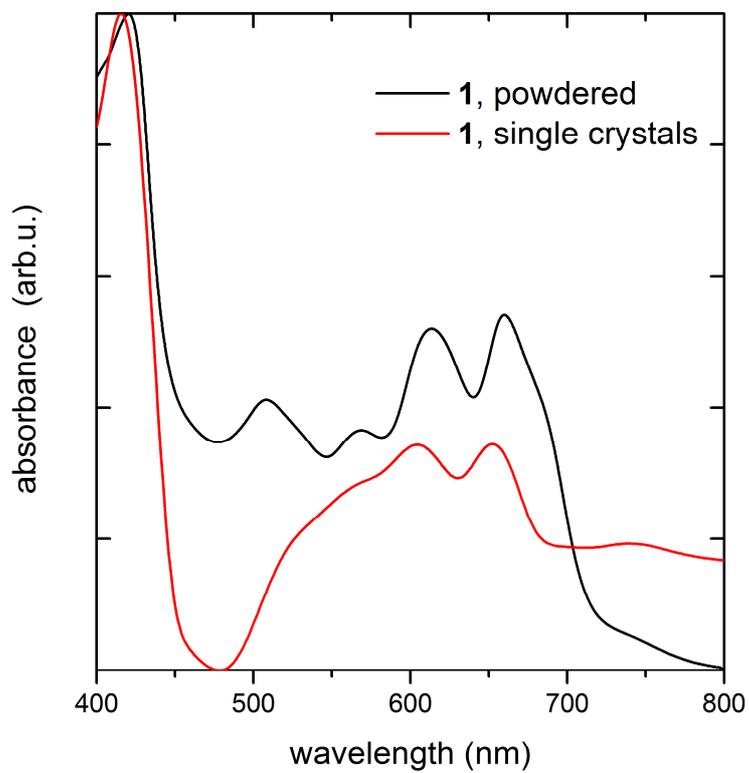

**Figure S6.** UV-Vis-spectra of single crystals and a finely powdered sample of **1**.

**X-ray Photoelectron Spectroscopy**

The survey spectrum of [H$_6$TPyP][BiCl$_6$]$_2$ reveals that both Bi and Cl are present within the surface region of the sample. The spectra were recorded with monochromatic AlK$\alpha$ radiation (1486 eV) under normal emission.

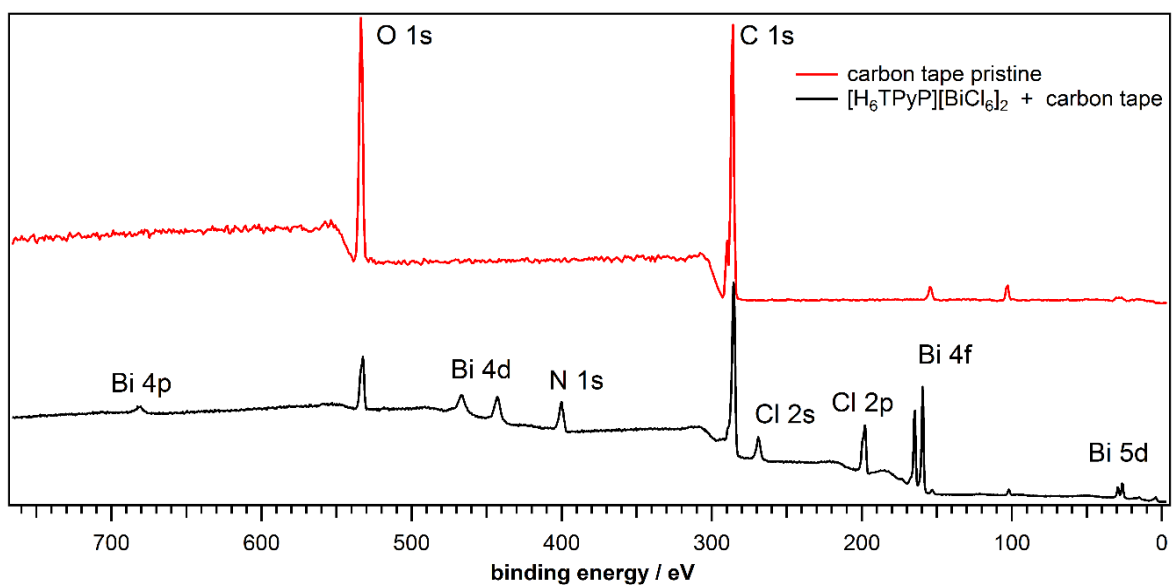

**Figure S7.** Survey spectra of the utilized carbon tape and [H$_6$TPyP][BiCl$_6$]$_2$ sticking to the carbon tape.